\documentclass{aastex631}
\usepackage{url}

\begin{document}

\title{New Post-DART Collision Period for the Didymos System: Evidence for Anomalous Orbital Decay}
\author[0009-0003-1106-0447]{Taylor Gudebski}
\affiliation{Thacher School, 5025 Thacher Rd., Ojai, CA 93203, USA}

\author[0009-0001-2900-7834]{Elisabeth Heldridge}
\affiliation{Thacher School, 5025 Thacher Rd., Ojai, CA 93203, USA}

\author[0009-0000-5029-3700]{Brady McGawn}
\affiliation{Thacher School, 5025 Thacher Rd., Ojai, CA 93203, USA}

\author[0009-0009-5189-1003]{Elle O Hill}
\affiliation{Thacher School, 5025 Thacher Rd., Ojai, CA 93203, USA}
\affiliation{University of California at Berkeley, Berkeley, CA 94720, USA}

\correspondingauthor{J. Swift}
\email{jswift@thacher.org}
\author[0000-0002-9486-818X]{Jonathan J. Swift}
\affiliation{Thacher School, 5025 Thacher Rd., Ojai, CA 93203, USA}

\author[0000-0002-2093-6960]{Henry Zhou}
\affiliation{Thacher School, 5025 Thacher Rd., Ojai, CA 93203, USA}

\begin{abstract}
On September 26, 2022, NASA’s DART spacecraft impacted Dimorphos, the secondary asteroid in the (65803) Didymos system, so that the efficiency with which a satellite could divert an asteroid could be measured from the change in the system’s period. We present new data from the Thacher Observatory and measure a change in period, $\Delta P = -34.2 \pm 0.1$\,min, which deviates from previous measurements by $3.5\,\sigma$. This suggests that the system period may have decreased by $\sim 1$ minute in the 20 to 30 days between previous measurements and our measurements. We find that no mechanism previously presented for this system can account for this large of a period change, and drag from impact ejecta is an unlikely explanation. Further observations of the (65803) Didymos system are needed to both confirm our result and to further understand this system post impact.
\end{abstract}


\section{Introduction} \label{sec:intro}
The near-Earth asteroid (65803) Didymos is a binary asteroid system and the subject of NASA's Double Asteroid Redirection Test (DART) Mission as well as ESA's Hera Mission. Previous characterization of the system \citep[see][and references therein]{Scheirich2022} reveals the primary asteroid, henceforth Didymos, to have a diameter of about 780\,m, an estimated total mass of $5.4\times10^{11}$\,kg, and a rotational period of 2.26 hours. Dimorphos, the secondary body and target of the DART satellite collision, has a diameter of about 170\,m, an estimated mass of about $4$-$5\times10^9$\,kg, and is expected to be synchronously rotating. The orbital period of Dimorphos at the time of impact was calculated to be $11.92147\pm0.00004$ hours; however, the period of Dimorphos was observed to be slowly changing before the DART impact. The measured quadratic drift in the mean anomaly pre-impact, $\Delta M_d = 0.15\pm0.047$\,deg\,yr$^{-2}$, corresponds to $\dot{P} = (1.5 \pm 0.4)\times10^{-9}$. The eccentricity of the mutual orbit is low, $\lesssim0.03$, as is the inclination of Dimorphos's orbit to the primary’s equator, $\lesssim3^\circ$.

The DART spacecraft successfully impacted Dimorphos on 2022 September 26 at 23:14:24.183 UTC ($t_{imp} = \textrm{JD}\, 2459849.46834$), and a change in period of $\Delta P = -33.0\pm0.3$\,min was measured \citep{Thomas2023}. This period change can be attributed to the inelastic collision between the spacecraft and Dimorphos with additional momentum transferred to Dimorphos from the launching of ejecta.


\begin{figure}[hb!]
\plotone{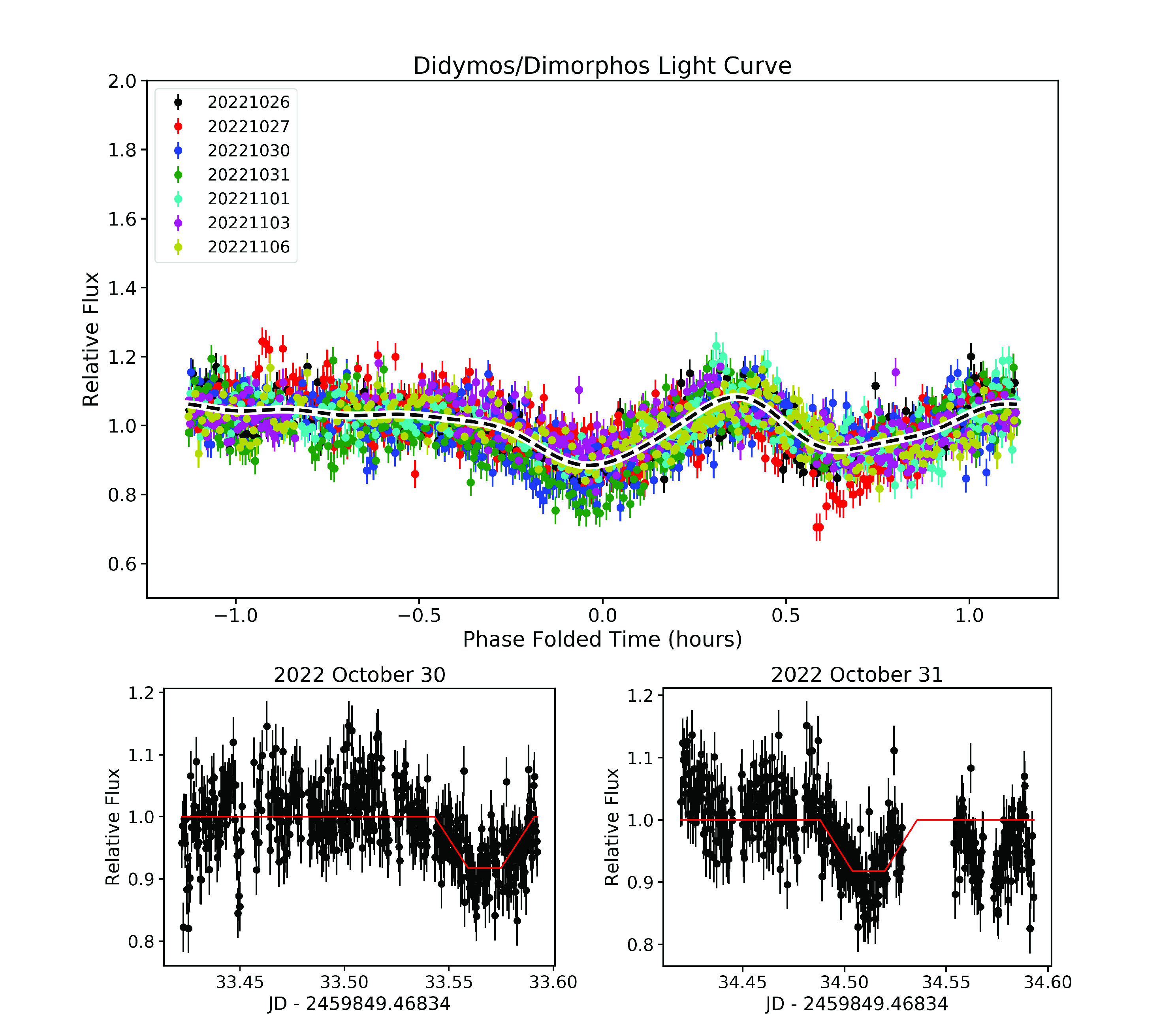}
\caption{(\textit{top panel}) Phase folded photometric measurements of (65803) Didymos used to derive a Fourier model of the rotational modulation to the light curve (dashed black line). (\textit{bottom panels}) Light curves of (65803) Didymos from 2022 October 30 and 31 corrected for rotational modulation  and showing the best fit primary eclipse model using our derived $\Delta P = -34.2$\,min. \label{fig:composite}}
\end{figure}

\section{Observations and Data Reduction} \label{sec:observations}
We observed the Didymos system in $r^\prime$ with the 0.7\,m telescope at the Thacher Observatory \citep{Swift2022} over 10 nights between UT 2022 September 15 and UT 2022 November 06 using non-sidereal tracking and integration times of 15 seconds to mitigate the streaking of background stars.
With readout and pointing corrections between each observation, the duty cycle of our measurements was 33 seconds. 

For each night of observations, we identified bright reference stars that remained in our $20\farcm8\times20\farcm8$ field of view for the entirety of the observations. Image calibrations and differential aperture photometry were performed in \texttt{python} using tools available in \texttt{astropy}. All light curves were phase folded to the 2.26 hour rotation period of Didymos with the zero phase chosen to correspond to the deepest dip in the rotational modulation. Seven nights of data were used to perform a Fourier decomposition of the phase folded light curves using 7 terms. The phase folded data and Fourier model can be seen in the top panel of Figure~\ref{fig:composite}.

The Fourier model was then divided into the data to reveal dips that could be attributed to mutual events (eclipses and occultations). 
Observations from two consecutive nights, UT 2022 October 30 and 31, revealed one $\sim 10$\% dip each that we attributed to primary eclipse events and are shown in the bottom panels of Figure~\ref{fig:composite}.

\section{Results and Analysis} \label{sec:results}
To estimate a change in period from the DART impact consistent with the two candidate mutual events in our data we used the geocentric predictions of \cite{Scheirich2022}\footnote{\url{https://www.asu.cas.cz/~asteroid/Didymos_2022-2023_events_update2022July.txt}} which extend beyond the impact time and account for the quadratic mean motion drift, but do not attempt to correct for a period change due to the impact. For each type of mutual event, we fit a linear ephemeris which we then subtract from the predictions to obtain a time correction due to the phase angle between the Earth, Didymos, and the Sun as a function of Julian Day. We then apply a new system period starting at the time of impact, $t_{imp}$, that produces a preliminary prediction for mutual event times that is time corrected for phase angle using a cubic spline interpolation of the linear fit residuals.

To measure the period change of Didymos due to the DART impact, we first modeled the two dip shapes with a single trapezoidal shape to obtain optimal parameters and then constructed $\chi^2$ as a function of $\Delta P$. We then directly evaluated $\chi^2$ in steps of 0.02 seconds for $-35 < \Delta P < -32$\,min to find a best fit period change of $\Delta P = -34.2 \pm 0.1$\,min. This corresponds to a $3.5\,\sigma$ discrepancy with $\Delta P = -33.0 \pm 0.3$\,min measured by \cite{Thomas2023}.

\section{Discussion} \label{sec:discussion}
The data we used to measure the post-collision period change was taken about 20-30 days after the data used by \cite{Thomas2023}, and we interpret our results to indicate that the system period may have shortened in this time. Before the DART collision, a quadratic drift in the mean anomaly of (65803) Didymos was found by \cite{Scheirich2022}. However, the corresponding period change of $\dot{P} = 1.5\times10^{-9}$ was accounted for in the mutual event predictions and is also 4 orders of magnitude too small to account for the difference we see. Therefore, whatever effect was causing the orbital decay before the collision cannot account for the discrepancy we observe; this includes the binary YORP effect, mutual tides, differential Yarkovsky force, nodal precession, and mass loss from the primary \citep{Trogolo2023}.

It has been estimated that between 0.9 and $5.2\times10^7$\,kg of ejecta were released in the impact \citep{Roth2023}. If some of the ejecta ended up in the orbital path of Dimorphos, this could cause Dimorphos's orbit to decay due to drag. An order of magnitude estimate for the mass of ejecta, $\Delta m$, needed to cause a period change from $P_i$ to $P_f$ is
\begin{equation}
    \Delta m = m_i\left[\left(\frac{P_i}{P_f}\right)^{1/3} -1 \right].
    \label{eq:dm}
\end{equation}
assuming that the total mass of the system does not change significantly. By this estimation, the amount of debris in the orbital path of Dimorphos needed to account for this period change would be $\sim 3\times10^6$\,kg, or $\sim 5$ to 30\% of the total ejecta mass. This seems unlikely given the estimated ejecta speeds, the low escape velocity of $\sim 0.2$\,m/s, and the short timescale for debris to be swept away by solar radiation pressure. 

\begin{acknowledgments}
We thank Konstantin Batygin and Jason Wright for helpful and encouraging conversations at the 242nd meeting of the American Astronomical Society in Albuquerque, NM.
\end{acknowledgments}

\vspace{5mm}
\facility{Thacher Observatory: 0.7m}

\software{\texttt{astropy} \citep{2018AJ....156..123A}, \texttt{scipy} \citep{2020SciPy-NMeth}, Source Extractor \citep{1996A&AS..117..393B}}

\bibliography{references}{}
\bibliographystyle{aasjournal}

\end{document}